\documentclass[conference]{IEEEtran}
\IEEEoverridecommandlockouts
\usepackage{cite}
\usepackage{amsmath,amssymb,amsfonts}
\usepackage{algorithmic}
\usepackage{graphicx}
\usepackage{textcomp}
\usepackage{xcolor}
\usepackage{listings}
\usepackage{nameref}
\def\BibTeX{{\rm B\kern-.05em{\sc i\kern-.025em b}\kern-.08em
    T\kern-.1667em\lower.7ex\hbox{E}\kern-.125emX}}
\begin{document}

\title{Escape the Fake: Introducing Simulated \\Container-Escapes for Honeypots}

\author{\IEEEauthorblockN{1\textsuperscript{st} Daniel Reti}
\IEEEauthorblockA{Intelligent Networks Research Group\\
German Research Center for Artificial Intelligence\\
67655 Kaiserslautern, Germany\\
Email: }
\and\IEEEauthorblockN{2\textsuperscript{nd} Norman Becker}
\IEEEauthorblockA{Intelligent Networks Research Group\\
German Research Center for Artificial Intelligence\\
67655 Kaiserslautern, Germany\\
Email: norman.becker@dfki.de}
}

\author{\IEEEauthorblockN{Daniel Reti\IEEEauthorrefmark{1},
Norman Becker\IEEEauthorrefmark{1}}
\IEEEauthorblockA{\IEEEauthorrefmark{1}Intelligent Networks Research Group, 
German Research Center for Artificial Intelligence, Kaiserslautern\\}
Email: {firstname}.{lastname}@dfki.de}

\maketitle

\begin{abstract}
In the field of network security, the concept of honeypots is well established in research as well as in production. Honeypots are used to imitate a legitimate target on the network and to raise an alert on any interaction. This does not only help learning about a breach, but also allows researchers to study the techniques of an attacker. With the rise of cloud computing, container-based virtualization gained popularity for application deployment. This paper investigates the possibilities of container-based honeypots and introduces the concept of simulating container escapes as a deception technique.
\end{abstract}


\section{Introduction}
Many defense-based protection strategies against attacks on IT infrastructure are well established and are proven to be effective. Firewalls, intrusion detection systems (IDS), network segmentation and strict authentication and update policies are at least expected on enterprise networks. Nonetheless, with these measures in place, there still might be penetrable gaps which will eventually be found by attackers. This means there is an asymmetry where the defender must be successful all the time while the attacker must only be successful once. This problem can be addressed using deception techniques, where false targets are presented aiming to expose attackers on interaction. Another problem deception addresses is the threat from malicious insiders.
The most common deception technique are honeypots which simulate a host on the network with the purpose to raise an alert on interaction while wasting the attacker's resources. 
Historically, honeypots are being deployed as bare-metal services or as virtual machines. With the ongoing infrastructure migration to the cloud, OS-level virtualization such as Linux Containers (LXC) or docker containers have gained momentum for several reasons. Containers offer an easy deployment of services similar to virtual machines, while being more lightweight, as multiple containers can share the same kernel and images. By this, container-based applications introduce a better scalability than virtual machines, as more container instances can be started quickly on demand.
Several works in academia as well as in the open-source community have leveraged the advantages of containers for honeypot use case. While most of these works focus on the easy of deployment for traditional honeypots \cite{Memari.14.04.201416.04.2014} \cite{SergiuEftimie.2016} \cite{Multari.2017}, the open-source project \textit{whaler}, simulates a vulnerable docker container and the honeypot \textit{Dockpot} starts containers on incoming SSH connections. 
In this work we present a novel docker honeypot concept which starts a new instance for each connection over an SSH proxy, and presents itself to the attacker as a vulnerable docker container. When the attacker attempts the escape, a successful escape is simulated, while remaining in the sandbox environment.
The contributions of this paper are the following:
\begin{itemize}
  \item Give an extensive review on the usage of containers for honeypots
  \item Introduce an SSH-proxy to create a new honeypot instance for each connection and log commands
  \item Introduce a docker honeypot which simulates a successful escape
  \item Discuss the implementation on a detailed level
\end{itemize}

The remainder of this work in structured as follows: In Chapter II virtualization in general and for deception are introduced. In Chapter III an attacker model is given and used to explain the benefits of the proposed concept. In Chapter IV chapter is explained how the proposed concept was implemented. In Chapter V the benefits and difficulties are discussed and in the conclusion as short summery of the work and the results is given.

\section{Background}
\subsection{Virtualization for Security}
In general, every replacement of physical resources with virtual replicas is referred to as virtualization. Traditionally virtual machines are used for the virtualization of computing infrastructure, but recently the much more lightweight OS-level virtualization has gained momentum, especially due to the popularity of cloud computing \cite{Gundall2020}. Reasons for virtualization are portability of applications, as the environment is already set up and can be deployed on every system, the serverless paradigm allows to bill only computational power that is being used, and the isolation which allows to strictly control the access to system resources and sandbox application enabling security and privacy when multiple tenants share a platform. In contrast to virtual machines, which provide a very high degree of isolation, in OS-level virtualization containers share the same kernel and therefore exploiting kernel vulnerabilities would allow lateral movement throughout container instances. In addition, it is a common weakness that wrongly configured containers allow leveraged privilege to access other containers on the same system. Such attacks are referred to as container escapes \cite{Combe.2016}

 \begin{figure}[htbp]
\centerline{\includegraphics[width=\columnwidth]{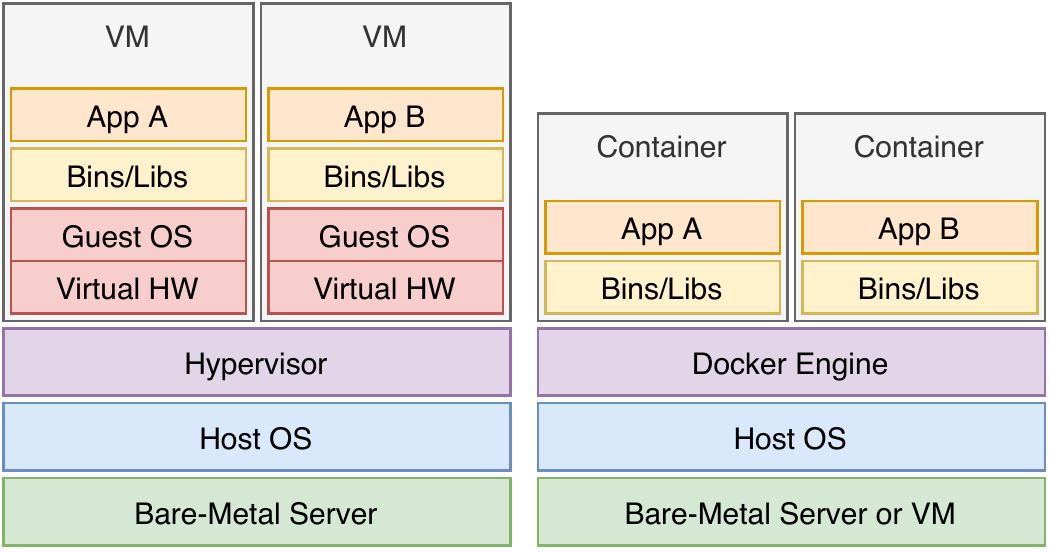}}
	\caption{Comparison of abstraction layers between VM and containers.}
\label{fig:VM}
\end{figure}

As shown in Figure \ref{fig:VM}, containers have a much lower overhead for applications, as no hypervisor is needed like in virtual machines. Instead, as container manager such as the docker engine, allows to create different container instances on the same host that share the same kernel, while they kernel based features provide the isolation and resource restriction needed. These features are mainly the namespaces and the cgroups feature. With namespaces processes may have a individual local instance of kernel objects and file system paths. With cgroups the access to system resources like CPU time, memory access, networking capabilities, disk I/O and processes is restricted. Another reason for the efficiency of containers is the overlay filesystem. With it multiple containers will share the same base image filesystem and only write modifications
This way multiple containers can share the underlaying image as it remains unchanged which saves storage and allows a faster deployment.

Container Escapes

\subsection{Virtualization Based Deception}
Leveraging virtual machines for honeypot use cases is a common and well-established technique due to the high level of isolation and easy deployment of VMs. With the rising popularity of cloud-infrastructures based the more lightweight OS-level virtualization instead of VMs, many honeypot techniques based on containers have been proposed by academia and the open source community. The following will first give an overview on container-based honeypots in academic literature and then present notable related open source projects. 

\subsubsection{Academic literature}
In 2014, Memari et al.. proposed a honeynet, a network of honeypots, based on LXC, focusing on the scalability of the honeynet on demand, suggesting that 7000 systems can be emulated at a time \cite{Memari.14.04.201416.04.2014}.
In 2015 Samri proposed the usage of container based honeypots to create interactive sessions using LXC, aiming to detect password-cracking attacks \cite{Smari.2015}.
In 2016 Eftimie et al. suggest using container based honeypots due to ease of deployment \cite{SergiuEftimie.2016}.
In 2017, Kerdowitsch et al. proposed LXC based honeypots aiming to minimize virtualization fingerprints \cite{Kedrowitsch.2017} and conducted experiments on the detectability of different virtualization technologies, concluding that containers are easily identifiable with existing tests \cite{Kedrowitsch.2017}.
In 2017 Mills et al. used container orchestration to deploy high-interaction honeypots and reset them after a compromise, while centrally collecting a stream of events from all honeypots \cite{Multari.2017}.
In 2018 Kyriakou et al. deployed well established honeypots using docker containers, pointing out the advantages of availability, resiliency and automatic deployment \cite{AndronikosKyriakou.2018}. 
In 2018 Sever et al.   analyzed the risk of applying container based honeypots\cite{Sever.21.05.201825.05.2018}. The main identified risk hereby was using improperly configured images. 
In 2019 Osman et al. propose a sandbox environment which mirrors the production environment and uses SDN to reroute suspicious connections into the sandbox environment. \cite{AmrOsman.2019}
In 2020 Bistarelli et al. studied log data of a publicly reachable instance of the docker honeypot \textit{whaler} to see of the deliberate vulnerability is being exploited. The result is that of 5 attackers who logged on and executed commands, one exploited the intended vulnerability  \cite{StefanoBistarelliEmanueleBosiminiFrancescoSantini.2020}.
In 2020 Sedlar et al. conducted a honeypot experiment by deploying three honeypots, namely \textit{Cowerie}, \textit{Diaonea}, \textit{HTTP trap} and \textit{Filebeat}, using Docker and Kubernetes \cite{Sedlar.07.07.202009.07.2020}.

\subsubsection{Open source projects}
\textbf{Honeytrap} is the most popular honeypot docker image on docker hub \cite{Honeytrap.2020} with more than 106 thousand downloads in November 2020. It is not a honeypot, but rather a framework for running, monitoring and managing honeypots.

\textbf{CommunityHoneyNetwork} is another open source honeypot manager which comes as a docker image \cite{JesseBowling.2020}. 

\textbf{Dockpot} is a docker based SSH honeypot which functions as an SSH proxy and forwards attackers into honeypots. It creates new container on connection and destroys the container when all connections are closed \cite{Aabed.2015}

\textbf{Dockerpot} is a docker based honeypot including open source honeypots such as Kippo, Glastopf, Dionaea and Thug. It creates a new container for each connection. The forwarding is implemented creating an iptables rule \cite{PeterKasza.2015}.

\textbf{Whaler} is a docker based honeypot which intends to be seen as an insecure docker daemon API. To achieve this, it capsules the insecure docker daemon in another docker container.  \cite{Oncyberblog.2018}
\\\break
Apart from the previously listed projects, many open source honeypots are also available on docker hub. For example, a search for the well-known honeypot \textit{cowrie} on docker hub retrieves 99 results (November 2020).

\section{Architecture}

 \begin{figure}[htbp]
\centerline{\includegraphics[width=\columnwidth]{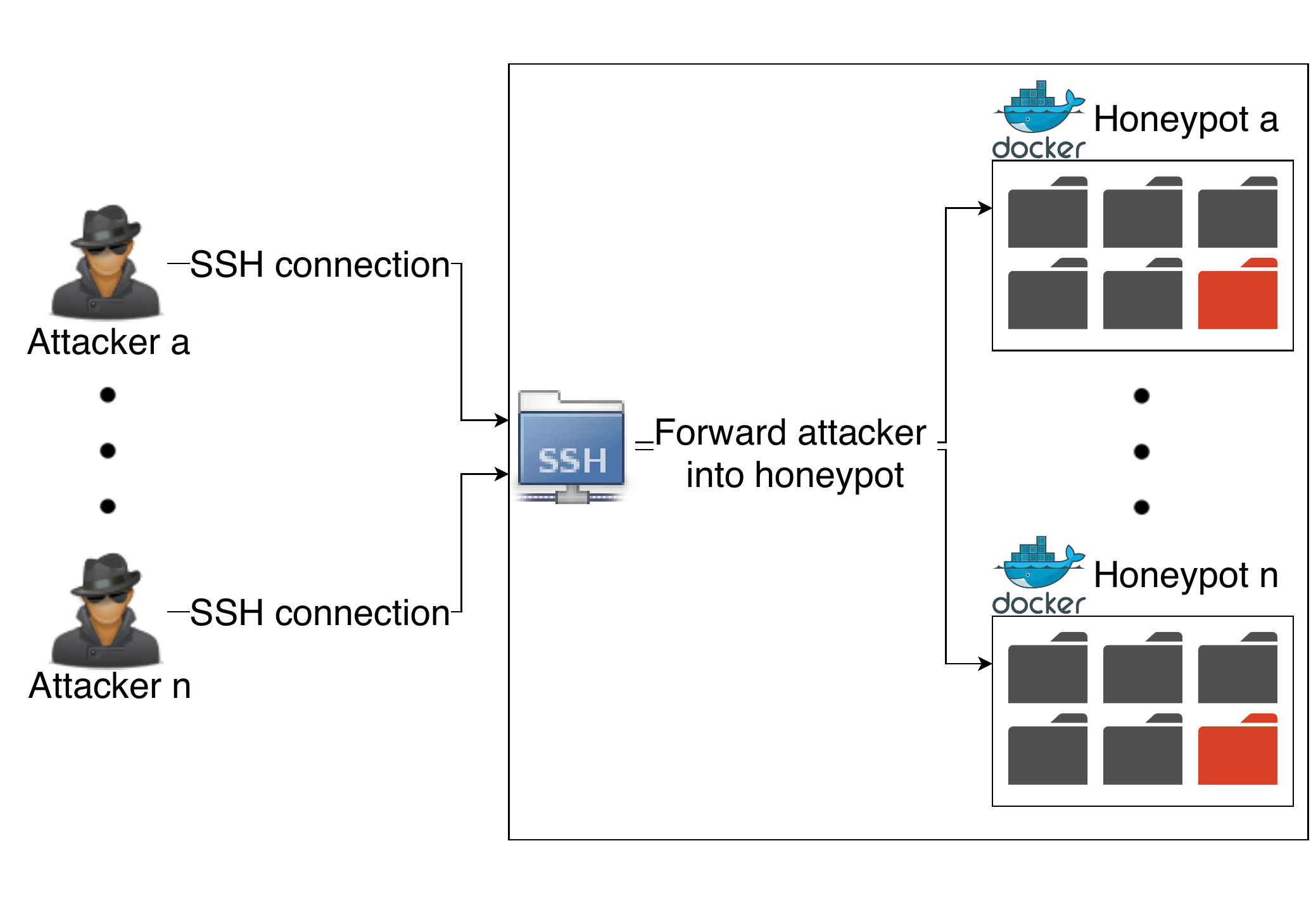}}
	\caption{Overview of the SSH-proxy. For each SSH connection a new honeypot instance is started.}
\label{fig:Overview}
\end{figure}

The architecture of the honeypot is composed of three aspects, the SSH-proxy, the simulated container escape and the fake host system.

\subsection{SSH-Proxy}
For any attacker to find the honeypot, the SSH port TCP/22 is open for incoming connections. On a connection attempt, the credentials will always be successfully on the first attempt. For each used combination of username and password, a new container instance will be started with a local account using the respective credentials, and the incoming connection will be forwarded into the container. All activity of the attacker is tracked over the SSH server.

\subsection{Simulated Container Escape}
A sophisticated attacker might recognize the container environment and attempt an escape. The first step of the simulated escape is finding the root credentials. These can be found in a text file called \textit{~/password.txt}. Having root permission in the container the attacker can execute \textit{mount}, allowing to mount the partition \textit{vda1}, gives access to the host file system. Allowing to change any file on the host and therefore escalating privileges to host and therefore escaping the container. Due to the SELinux configuration \textit{mount} is the only command the attacker is allowed to execute. 

\subsection{RCE on Fake Host}
From this point the attacker has multiple possibilities to gain Remote Code Execution (RCE). One simple method is given with a scheduled job in \textit{/etc/crontab} which will execute the file \textit{~/backup.py} every minute. The attacker can simply overwrite the content of \textit{backup.py} to execute arbitrary commands and gain shell access.

\section{Implementation}
The main components for the implementation of the honeypot, as shown in Figure \ref{fig:Overview}, are the SSH-proxy, the docker containers themselves, the container hardening and the mount command. 

The SSH-proxy plays an essential role, as it controls the SSH-login and manages the containers i.e. starts a new container for every connection and logs traces of an attacker. For the implementation of the SSH-proxy, Paramiko was used, a library for Python3, which allows to easily create and modify an SSH-sever. The Paramiko service listens on TCP port 22. The SSH-proxy acts as a reverse-proxy with the following behavior:
It accepts every arbitrary username and password and saves them. After this, a new docker container instance is created from the base image and inside the container a user is created with the SSH credentials used by the attacker. When the generation of the docker container is finished, the SSH connection of the attacker is forwarded into the container.

Each container is mapped to a user using a unique ID, so when an attacker reconnects, the connection will automatically be forwarded into to the respective belonging container. Commands run inside a container are transmitted over the SSH-proxy and can therefore be intercepted and written into a log file.

As previously mentioned, docker was used as the container software. Each docker container is created from a docker image. The image used for this work is based on Ubuntu, has SSH installed and a text file with the root password is created. Root privileges are needed to use the \textit{mount} command.

To granularly control the permissions on the container, hardening was used. As the attacker is granted root permissions, the container hardening is needed on the honeypot, to prevent further damage and only allow a controlled \textit{mount} escape. An improperly configured container hardening on a privileged container might cause severe harm to the underlying system and adjacent containers. Depending on the operating system, different ways to harden a container might apply. In this work \textit{SELinux} was used, as the host OS is Centos8, for Ubuntu \textit{Apparmor} would be preferred. 
\textit{SELinux}, in contrast to \textit{Apparmor}, denies every command until it is specifically allowed. To use \textit{SELinux}, it must be enabled for the docker daemon and the rule to whitelist the \textit{mount} command has to be applied. This is done with the following commands:
\noindent
\begin{lstlisting}[language=bash,
    stepnumber=1, breaklines=true,]{code} 
sudo auserach -c 'mount' --raw | audit2allow -M my-mount
\end{lstlisting}
followed by
\begin{lstlisting}[language=bash,
    stepnumber=1, breaklines=true,]{code} 
semodule -X 300 -i my-mount.pp
\end{lstlisting}
Because \textit{SELinux} is added to the docker daemon, a specific \textit{SELinux} security flag is not needed, only the flag '{\lstinline[language=bash]!--cap-add=SYS_ADMIN!}' must be added, in order to use mount inside a container. 

In \textit{Apparmor}, a profile could be created, which specifies to which binaries access is denied. So for this work the \textit{Apparmor} profile "aa-self" is created, which allows the use of \textit{mount} but denies the \textit{cgroups} access, which could be used for an escape without \textit{mount}. To apply the rule, the \textit{Apparmor} profile needs to be uploaded into the kernel.
This could be done by
\begin{lstlisting}[language=bash,
    stepnumber=1, breaklines=true,]{code} 
apparmor_parser -r -W docker_honeypot/security/aa-self
\end{lstlisting}
At the generation of the docker container we need to set the flags 
\begin{lstlisting}[language=bash,
    stepnumber=1, breaklines=true,]{code} 
--cap-add=SYS_ADMIN --security-opt apparmor=aa-self
\end{lstlisting}
With the flags "--memory=512m" and  "--cpus=.5", memory and CPU time are limited to protect system resources from being used for a botnet. 

With \textit{mount} available in a container, a common attack method is to mount the partition of the host system, typically named \textit{sda1} or \textit{vda1}. When mounted, an attacker is in position to manipulate the host file system and therefore escape the Docker containter. To simulate this process, the partition names were changed. A second partition is created, named \textit{sda2}. Then the names inside \textit{/dev} were changed from \textit{sda1} to \textit{tty66} and from \textit{sda2} to \textit{sda1}. This can be done with the \textit{mv} command. Thereby the goal is to obfuscate the real partition, as no partition should be expected under \textit{tty66}. As a result, there are two partitions: \textit{sda1} as the fake partition and \textit{tty66} with the real boot partition. The system still works as normal and \textit{sda1} is still listed at the mount command. When \textit{sda1} partition is attempted to mount, the fake partition is mounted. This fake partition is shown in Fig.2 marked in red. 

A disadvantage of this is, that a partition can just depict files and not a real system with services. To address this another docker container is used, which poses as a fake host system. To get a connection between the partition and the docker container, the container shares a volume with the partition. 

To allow the attacker to execute remote commands, one strategy is to write a scheduled job. As the cronjobs themselves are not writable for the attacker, a prepared cronjob is placed which will execute a \textit{python3} script inside the volume folder named \textit{backup.py}. This is a common misconfiguration that attackers should expect. The script is executed inside the fake host container by the cronjob every minute. The attacker can modify the script to have a remote shell sent from the fake system.
 \begin{figure}[htbp]
\centerline{\includegraphics[width=\columnwidth]{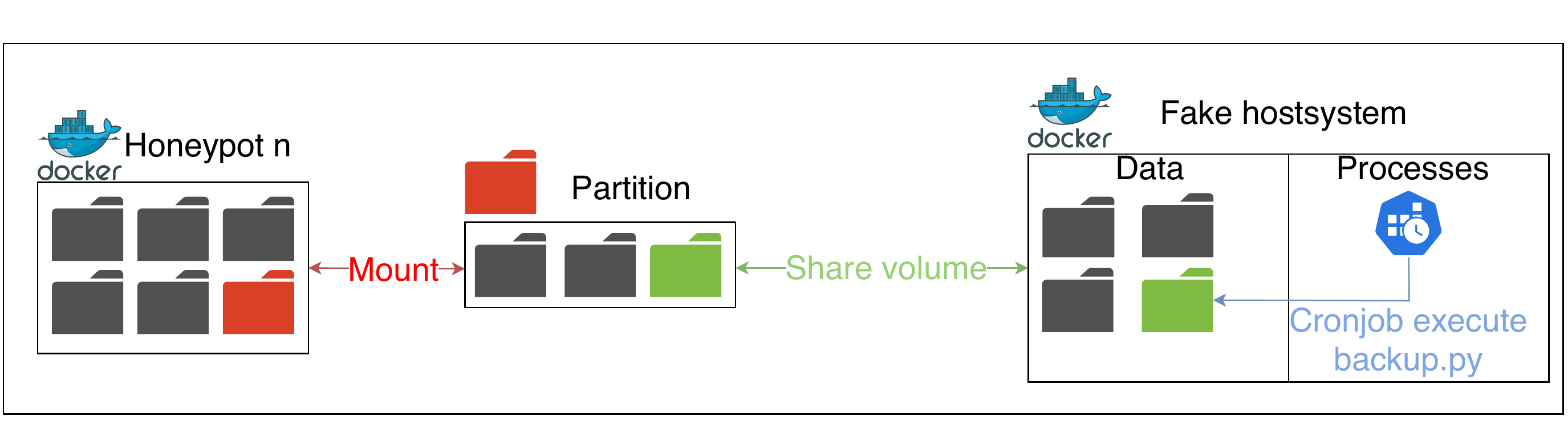}}
	\caption{Overview of the simulated container escape. A seemingly real host partition could be mounted. In reality a fake host system uses this partition as a share volume.}
\label{fig:Escape}
\end{figure}

This only works assuming the attacker cannot modify any system files. Since the attacker can modify every file inside the partition, another way to escape would be to create an own cronjob, create a user by modifying \textit{/etc/passwd} or add SSH keys. To avoid inconsistencies between partition and fake system, bind mounts are added to the docker container. This can be achieved using the command
\begin{lstlisting}[language=bash,
    stepnumber=1, breaklines=true,]{code} 
--mount type=bind,source=/home/Desktop/m1/etc,target=/etc
\end{lstlisting}to bind \textit{/etc}. 
All escape methods have in common that the command logging by the SSH-Proxy will not be possible, since another connection is established. Therefore, the logs need to be traced inside this container.
For this all shells except for bash are blocked. Bashes can be set to log every executed command. The log files inside the docker container are set to read-only for none root users. Since the attacker will not be able to get root access on this container, the logs cannot be deleted. One the honeypot the logs would not be protected since there the attacker can get root privilege and delete the bash logs. 
Finally, it should seem for the attacker that the escape has been successful. All further actions can be monitored.

\section{Discussion}
Traditional honeypots increase the chance for attackers to be detected when searching for targets. When a regular honeypot is deployed in form of a docker container, an attacker might recognize the honeypot and attempt an escape. The proposed system offers the same advantages of a traditional honeypot, but also increases the uncertainty for attackers when approaching a container escape. One issue which needs to be addressed in the proposed system faces is, that an attacker might attempt to avoid the command logging over the SSH proxy, by establishing a new shell connection over a remote shell. Another
Problem might be the privileged access to the Docker in Docker. The reals host partition is hidden but might be found and mounted, so a successful escape is possible. This could be addressed by running the honeypot in a safe environment, like a virtual machine. Further research is needed to evaluate the concept. A field study is needed to quantify the success rate of the approach.

\section{Conclusion}
This work compared different approaches of virtualization applications for honeypots and proposed a container-based honeypot. The proposed honeypot distinguishes itself from other projects through two features. First, it uses an SSH-Proxy to start a new container instance for each attacker and log all commands and second it simulates a successful container escape. The latter is a counter measure to a scenario where an attacker might detect the honeypot of being a docker container and attempt an escape. This work gave details of how this concept was implemented. It shows to be technically working and appears to be a promising concept. A field study of the proposed system is yet to be done.

\section{Acknowledgement}
This research was supported by the German Federal Ministry of Education and Research (BMBF) within the IUNO Insec project under grant number 16KIS0932, and within the SCRATCh project under grant number 01IS18062E. The SCRATCh project is part of the ITEA 3 cluster of the European research program EUREKA. The responsibility for this publication lies with the authors.
\bibliographystyle{IEEEtran}
\bibliography{literature}
\vspace{12pt}

\end{document}